# Coupling Onset of Cyclone Upward and Rotation Flows in a Little Bottle

Shigeo Kawata[1]

A coupling onset of the cyclone upward and rotation flows is experimentally demonstrated in a little bottle. The rotating flow provides a pressure increase in the outer part of the rotating flow by its centrifugal force. When a gradient of the fluid rotation appears along the rotation axis, the higher-pressure area is localized and pushes the fluid in a low pressure. Then the fluid staying in the central area of the rotation is pushed up along the rotation axis, and the upward wind is enhanced. In this coupling mechanism the rotation gradient is the key; the coupling of the rotation and the upward fluid flow is essentially important for a cyclone buildup, and is well explained experimentally and theoretically.

Hurricanes or tropical cyclones or volcanic cyclones have brought us serious damages, and underlying physics has been studied extensively [1-6]. On the other hand, a small size of forced cyclone is used for vacuum cleaner to separate dust particles from clean air [7, 8]. In ocean warm eddies influences ocean current pathway and sea level, and may show an anomalous tide[8,9]. Even in the recent tsunami and earthquake disaster[10] in the east Japan in March 11, 2011, ocean eddies also appeared. The physical mechanism for the cyclone or eddy initiation is not still perfectly clarified[1-9,11]. Cyclone is associated with a fluid rotation and an upward flow. We present one of the mechanisms for the cyclone initiation using two stratified fluids in a little bottle

---

[1]Dept. of Advanced Interdisciplinary Sciences, Utsunomiya University, Yohtoh 7-1-2, Utsunomiya 321-8585, Japa

experimentally as shown in Fig. 1. The relating physics of the coupling of the rotation and the upward flow was discovered by the experiment and by a theoretical estimation.

Figure 1 shows our simulation experiment for an upward flow generated by a rotating flow. In a little cylindrical bottle a sticky oil and a water layer are stratified. The two fluids have the same mass density for the present research purpose in order to avoid the Rayleigh-Taylor instability [12,13], which may complicate the present phenomenon. The bottle is initially stationary, and then the bottle starts its rotation about its axis. The upper oil layer has a nonslip boundary condition at the bottle wall, though the water slips at least at the initial period of the bottle rotation. Therefore, a rotation gradient is created in the bottle. The centrifugal force acts only on the oil, and the centrifugal force pushes the oil outward. A high pressure area is generated, and a part of water is pushed downward and finally inward in the bottle. The oil and water circulation creates an upward flow around the axis.

Figures 2 show experimental results for a coupling mechanism of the fluid rotation and the upward flow at **a** $t$=0s, **b** 0.2s, **c** 0.4s, **d** 0.6s and **e** 0.8s. The upper layer is an oil, whose mass density is controlled to the same value of the water. The lower water layer is colored by a black ink. The bottle is set to the center of a turntable, rotating with 2.5rps. The little bottle is initially stationary, and then the bottle starts its rotation about the rotation axis at $t$=0s. The upper sticky oil layer has the nonslip boundary condition at the bottle wall, and the water slips at the initial period of the bottle rotation. Therefore, a rotation gradient appears in the bottle clearly. The transparent upper oil layer has a high pressure near the bottle wall, and the oil pushes down in this case the black water. The outer water layer moving downward hits the bottom of the bottle, and the central part of the black water is forced to move upward (see Figs. 2**b**-**e**). The bottle inner radius is about 1.9cm, and the rotation is 2.5rps. So the outer high pressure is estimated by $p = p_0 + \rho r^2 \omega^2 /2$.[14] Here $p$ is the pressure at the radius $r$ in the oil layer, $p_0$ is the pressure at the rotation center of $r$=0, $\omega$ the rotation angular speed and $\rho$ the mass density (~1g/cm$^3$). The pressure increase $\Delta p = p - p_0$ is estimated by $\Delta p = \rho r^2 \omega^2 /2$. Here $r$~1.9cm and $\omega = 2\pi \times 2.5$. $\Delta p$~44.5Pa. The black water should move upward by ~4.45cm in the bottle. In Fig. 2e the black water moves up to ~4.5cm, and the results shown in Figs. 2 is well supported by the theoretical estimation.

Shown in Fig. 3 is a coupling mechanism of the rotation flow and the upward wind associated with cyclone. In a cyclone upward flow may be initially generated at the heated ground or sea, and the rotation flow is created by the Coriolis force[15]. When the upward flow is enhanced by the vertical gradient of the rotation speed as demonstrated above, the cyclone strength is boosted up and it may bring huge damages for our society and life[1]. The circulation enhances the cyclone / eddy power. After reaching an equilibrium state of the cyclone, the cyclone structure may be described approximately by the vortex equation[14]: $\frac{\partial \vec{\Omega}}{\partial t} = rot(\vec{v} \times \vec{\Omega})$. Here $\vec{v}$ is the fluid velocity and $\vec{\Omega} = rot\,\vec{v}$ the vortex. In the equilibrium state, $\frac{\partial \vec{\Omega}}{\partial t} = 0$ and one of the solutions for the equilibrium state must satisfy a force-free configuration[16-18]: $\vec{v} = \lambda \vec{\Omega}$. Then we obtain the following equation: $\Delta \vec{v} + \frac{\vec{v}}{\lambda^2} = 0$. In the cylindrical coordinate $(r,\theta,z)$, the following explicit solution is obtained: $v_r = A_r J_1(\beta r)\sin(kz)$, $v_\theta = A_\theta J_1(\beta r)\cos(kz)$, $v_z = A_r \frac{\beta}{k} J_0(\beta r)\cos(kz)$. Here $\beta^2 = \frac{1}{\lambda^2} - k^2$, $\lambda$ is the scale length in the radial direction of the cyclone, $\lambda_z = \frac{2\pi}{k}$ the scale length in the vertical direction ($z$), and in the cyclone phenomenon $\frac{1}{\lambda^2} >> k^2$. The solution is shown in Fig. 4 for the cyclone equilibrium state. Fig. 4a shows the $(v_r, v_z)$ distribution, b the $v_\theta$ distribution at $z$=0 and c a typical test particle trajectory in the cyclone. In the example, $A_r = 1.0$, $A_\theta = 5.0$, $\lambda = 1.0$ and $\lambda_z = 100.0$.

Based on our experiment and considerations above, the coupling onset mechanism is clearly explained. The fluid circulation physical mechanism comes from the rotation gradient in the vertical direction in the case concerned in the letter.

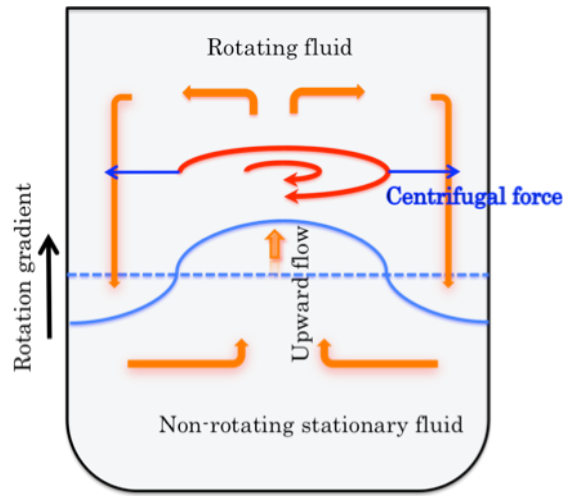

Fig. 1 Simulation experiment for an onset of upward flow in a cyclone. In a little cylindrical bottle a sticky oil and a water layer are stratified. The two fluids have the same mass density for the present research purpose in order to avoid the Rayleigh-Taylor instability, which may complicate the present phenomenon. The bottle is initially stationary, and then the bottle starts its rotation about its axis. The upper oil layer has a nonslip boundary condition at the bottle wall, though the water slips at least at the initial period of the bottle rotation. Therefore, a rotation gradient is created in the bottle. The centrifugal force acts only on the oil, and the centrifugal force pushes the oil outward. A higher pressure area is generated and a part of water is pushed downward and finally inward in the bottle. The oil and water circulation creates an upward flow around the axis.

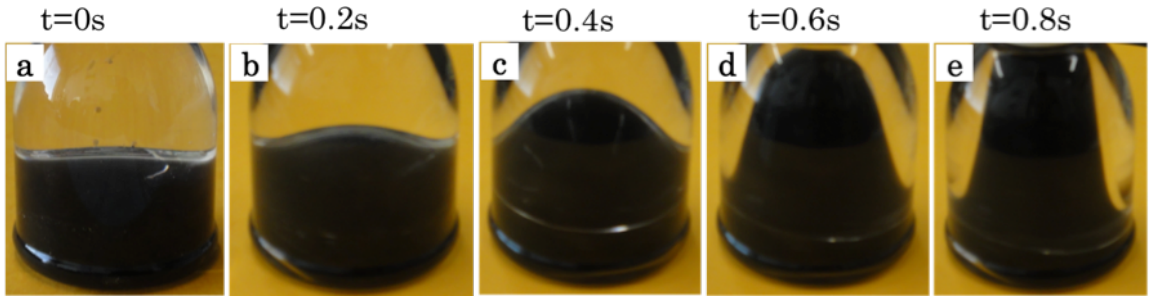

Fig. 2 Simulation experiment for an onset mechanism of upward wind associated with cyclone at **a** *t*=0s, **b** 0.2s, **c** 0.4s, **d** 0.6s and **e** 0.8s. The upper layer is an oil, whose mass density is controlled to the same value of the water. The lower water layer is colored by a black ink. The bottle is set to the center of a turntable, rotating with 2.5rps. The little bottle is initially stationary, and then the bottle starts its rotation about its axis. The upper oil layer is sticky and has a nonslip boundary condition at the bottle wall, though the water slips at least at the initial period of the bottle rotation. Therefore, a rotation gradient is created in the bottle. The centrifugal force acts only on the oil, and the centrifugal force pushes the oil outward. A higher pressure area is generated and a part of water is pushed downward and finally inward in the bottle. The upward flow onset is clearly presented in the bottle.

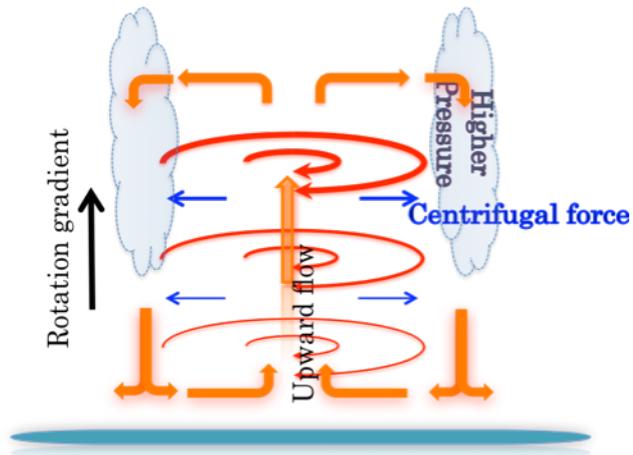

Fig. 3 A coupling onset mechanism between the upward wind and the rotation flow associated with cyclone. In cyclone or hurricane there may be a rotation gradient. The centrifugal force acts only on the rotating fluid, and the centrifugal force pushes the rotating fluid outward. A higher pressure area is generated and a part of non-rotating fluid is pushed downward and inward as shown at the bottom ground, for example. Around the central axis the upward flow is created. The coupling of the upward flow and the rotation may enhance the cyclone power.

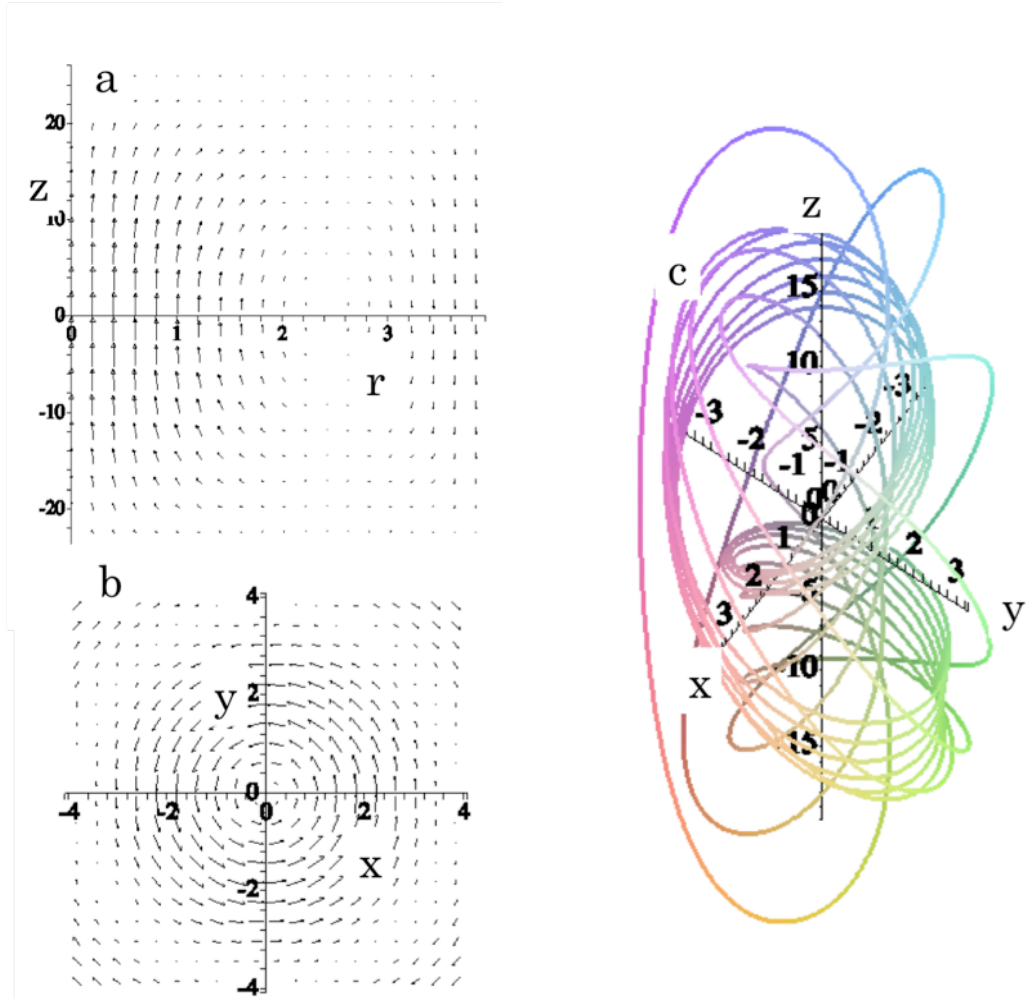

Fig. 4 An example analytical solution for the cyclone equilibrium state. **a** $(v_r, v_z)$ distribution, **b** $v_\theta$ distribution at $z=0$ and **c** a typical test particle trajectory in the cyclone.

Supporting Online Material:　An demonstration experiment video is available.

**Acknowledgements** I acknowledge support through MEXT (Ministry of Education, Culture, Sports, Science and Technology, Japan), JSPS (Japan Society for promotion of science, Japan) and CORE (Center for Optical Res. And Education, Utsunomiya univ., Japan).

**Author Information** kwt@cc.utsunomiya-u.ac.jp